 \documentclass[10pt,preprint]{aastex}  %compact preprint style (MJA)
 \usepackage{natbib}
\def\ang{\AA}
\def\arcsec{\hbox{$^{\prime\prime}$}}

\def\gapprox{\lower.4ex\hbox{$\;\buildrel >\over{\scriptstyle\sim}\;$}}
\def\lapprox{\lower.4ex\hbox{$\;\buildrel <\over{\scriptstyle\sim}\;$}}

\shortauthors{ASCHWANDEN}
\shorttitle{Coronal Loop Twisting and Braiding}

\begin{document}

\title{		Helical Twisting Number and Braiding Linkage Number
		of Solar Coronal Loops} 

\author{        Markus J. Aschwanden$^1$}
\affil{         $^1)$ Lockheed Martin,
                Solar and Astrophysics Laboratory,
                Org. A021S, Bldg.~252, 3251 Hanover St.,
                Palo Alto, CA 94304, USA;
                e-mail: aschwanden@lmsal.com }

\begin{abstract}
Coronal loops in active regions are often characterized by
quasi-circular and helically twisted (sigmoidal) geometries, 
which are consistent with dipolar potential field models in 
the former case, and with nonlinear force-free field models 
with vertical currents in the latter case. Alternatively, 
Parker-type nanoflare models of the solar corona hypothesize 
that a braiding mechanism operates between unresolved loop 
strands, which is a more complex topological model. In this 
study we use the vertical-current approximation of a nonpotential
magnetic field solution (that fulfills the divergence-free and 
force-free conditions) to characterize the number of helical
turns $N_{twist}$ in twisted coronal loops. We measure the 
helical twist in 15 active regions observed with AIA and HMI/SDO 
and find a mean nonpotentiality angle (between the potential and 
nonpotential field directions) of $\mu_{NP} = 15^\circ \pm 
3^\circ$. The resulting mean rotational twist angle is 
$\varphi = 49^\circ \pm 11^\circ$, which corresponds to 
$N_{twist}=\varphi/360^\circ = 0.14\pm0.03$ turns with 
respect to the untwisted potential field, with an absolute
upper limit of $N_{twist} \lapprox 0.5$, which is far below the kink 
instability limit of $|N_{twist}| \gapprox 1$.  The number of 
twist turns $N_{twist}$ corresponds to the Gauss linkage 
number $N_{link}$ in braiding topologies. We conclude that 
any braided topology (with $|N_{link}| \ge 1$) cannot explain 
the observed stability of loops in a force-free corona, nor 
the observed low twist number. Parker-type nanoflaring can 
thus occur in non-forcefree environments only, such as in the 
chromosphere and transition region.
\end{abstract}


\section{	INTRODUCTION					}

In this study we are measuring physical 
parameters of coronal loops that are useful as observational 
constraints for testing of theoretical coronal heating models,
such as the observed helical twist number of loops. 
These measurements should clarify whether the magnetic
field in the solar corona exhibits sufficient helical twisting  
so that it can be described by a braiding topology, which plays 
a key role in Parker-type nanoflare-driven heating scenarios 
(Parker 1988). 

Observations of the coronal magnetic field that have been 
interpreted in terms of helically twisted or braided topologies 
include flux ropes 
(Rust and Kumar 1996; Gary and Moore 2004; Raouafi 2009), 
coronal loops (T\"or\"ok and Kliem 2003; 
T\"or\"ok et al.~2010; Prior and Berger 2012), 
spiral structures in the transition region (Huang et al.~2018),  
chromospheric tornadoes (Wedemeyer-B\"ohm et al.~2012),
emerging active regions (Liu and Schuck 2012), 
coronal mass ejections (Kumar et al.~2012),
eruptive filaments/prominences (Kumar et al.~2010; Koleva et al.~2012),
confined eruptions (Hassanin and Kliem 2016), and 
possibly braided finestructure in Hi-C observations
(Winebarger et al.~2013, 2014; Cirtain et al.~2013; 
Tiwari et al.~2014; Pant et al.~2015). 

First clear evidence for helically kinked magnetic flux ropes 
was found by Rust and Kumar (1996), based on the sigmoid
(S-shaped or Z-shaped) geometry of 103 transient loop structures observed 
in soft X-rays by Yohkoh. Evidence for a multiple-turn
helical magnetic flux tube that was erupting during a flare was 
suggested by Gary and Moore (2004), in a singular case. 
Also, a double-turn geometry in an erupting filament was suggested 
by Kumar et al.~(2010) in another single case. 

However, direct measurements of braiding parameters (such as
the number of helical twisting turns or the Gauss linkage number) 
have only been attempted in very few single cases 
(Portier-Fozzani et al.~2001; Malanushenko et al.~2011, 2012;
Thalmann et al.~2014). The measured helical twist is a very 
important quantity, because a magnetic topology can only be
be diagnosed as a braiding mechanism when it produces
multiple helical turns, which dis-qualifies phenomena that
exhibit a fraction of a single helical turn only.
In this Paper we determine the helical twist number in
active regions, using a newly developed nonlinear force-free
field code that is parameterized in terms of the helical
twist of unipolar magnetic charges, the so-called 
{\sl vertical-current approximation nonlinear force-free field 
(VCA-NLFFF)} code. A similar concept was previously applied using a 
Grad-Rubin-type NLFFF code (Malanushenko et al.~2011, 2012), 
and the Wiegelmann-NLFFF code (Thalmann et al.~2014). 

The content of this Paper contains a brief theoretical description
of the VCA-NLFFF code (Section 2.1), the braiding topology
and linkage number (Section 2.2), and the kink and torus 
instability (Section 2.3). Then we carry out data analysis
of AIA and HMI observations that is specifically designed
to measure the helical twist and the Gauss linkage number
(Section 3), followed by a discussion (Section 4) and 
conclusions (Section 5).

\section{	THEORY 						}

\subsection{	Nonlinear Force-Free Magnetic Field 		}

A physically valid coronal magnetic field solution has to satisfy
Maxwell's equations, which includes the divergence-freeness condition,
\begin{equation}
        \nabla \cdot {\bf B} = 0 \ ,
\end{equation}
and the force-freeness condition,
\begin{equation}
        \nabla \times {\bf B} = \alpha({\bf r}) {\bf B} \ ,
\end{equation}
where $\alpha({\bf r})$ represents a scalar function that depends
on the position ${\bf r}$, but is constant along a magnetic field
line. Three different types of magnetic fields are generally
considered for applications to the solar corona: (i) a {\sl potential
field (PF)} where the $\alpha$-parameter vanishes $(\alpha=0)$,
(ii) a {\sl linear force-free field (LFFF)} $(\alpha = const)$, and
(iii) a {\sl nonlinear force-free field (NLFFF)} with a spatially
varying $\alpha({\bf r}) \neq 0$. 

Due to the axi-symmetry of a divergence-free flux tube, it is useful 
to write the divergence-free condition (Eq.~1) and the force-free 
condition (Eq.~2) in spherical coordinates $(r, \theta, \varphi)$,
expressed with the magnetic field components ($B_r, B_\theta, B_\varphi)$,
where $r$ is the radial direction, $\theta$ the inclination angle,
and $\varphi$ the azimuth angle (Fig.~1a,b),
\begin{equation}
       (\nabla \cdot {\bf B}) = {1 \over r^2} {\partial \over \partial r}
        (r^2 B_r) 
        + {1 \over r \sin{\theta}} {\partial \over \partial \theta}
        (B_\theta \sin{\theta})
        + {1 \over r \sin{\theta}}
        {\partial B_\varphi \over \partial \varphi} = 0 \ ,
\end{equation}
\begin{equation}
        \left[ \nabla \times {\bf B} \right]_r =
        {1 \over r \sin{\theta}}
        \left[{\partial \over \partial \theta}
        (B_\varphi \sin{\theta}) -
        {\partial B_\varphi \over \partial \varphi} \right]
        = \alpha B_r \ ,
\end{equation}
\begin{equation}
        \left[ \nabla \times {\bf B} \right]_\theta =
        {1 \over r}
        \left[{1 \over \sin{\theta}} {\partial B_r \over \partial \varphi}
        -{\partial \over \partial r} (r B_\varphi) \right]
        = \alpha B_\theta \ ,
\end{equation}
\begin{equation}
        \left[ \nabla \times {\bf B} \right]_\varphi =
        {1 \over r}
        \left[{\partial \over \partial r}( r B_\theta )
        -{\partial B_r \over \partial \theta} \right]
        = \alpha B_\varphi \ .
\end{equation}

Due to the nonlinearity of the equation system, no general
analytical solution of the magnetic field ${\bf B}({\bf r})$ has
been obtained for the coupled equation system of Eqs.(1-2) or Eqs.(3-6).
However, an analytical approximation of a divergence-free and
force-free magnetic field solution has been derived for a vertical
current at the lower photospheric boundary, which twists a
field line into a helical shape (Aschwanden 2013a),
\begin{equation}
        B_r(r, \theta) = B_0 \left({d^2 \over r^2}\right)
        {1 \over (1 + b^2 r^2 \sin^2{\theta})} \ ,
\end{equation}
\begin{equation}
        B_\varphi(r, \theta) =
        B_0 \left({d^2 \over r^2}\right)
        {b r \sin{\theta} \over (1 + b^2 r^2 \sin^2{\theta})} \ ,
\end{equation}
\begin{equation}
        B_\theta(r, \theta) \approx 0 \ ,
\end{equation}
\begin{equation}
        \alpha(r, \theta) = {2 b \cos{\theta} \over
        (1 + b^2 r^2 \sin^2{\theta})} \approx 2 b  \ ,
\end{equation}
\begin{equation}
	b = {2 \pi N_{twist} \over L} = {\varphi \over L}\ ,
\end{equation}
where $B_0$ is the magnetic field strength vertically above a buried magnetic
charge at photospheric height, $d$ is the depth of the buried
magnetic charge, $L$ is the length of a magnetic field line,
and $b=\varphi/L$ is related to the azimuthal rotation angle $\varphi$,
from which the number $N_{twist}$ of helical turns can be calculated.
The accuracy of the analytical approximation can be verified by inserting
the solution of Eqs.~(7-10) into the equation system of Eqs.~(3-6),
which shows that all linear terms $\propto b r \sin \theta$ cancel out
exactly, so that only second-order terms $\propto b^2 r^2 \sin^2 \theta$
remain. Thus the approximation is most accurate for a near-potential
field solution with $|b r \sin \theta| \ll 1$. Note that the analytical 
solution of Eq.~(7-8) represents a generalization of the solution for
a straight (not divergence-free) flux tube with uniform twist (e.g.,
see textbooks by Priest 1982, 2014; Sturrock 1994; 
Boyd and Sanderson 2003; Aschwanden 2004).

We see that the nonpotential solution $(b \neq 0$)
degenerates to the potential field solution in the case of $(b=0)$,
\begin{equation}
        B_r(r) = B_0 \left( {d_j \over r_j} \right)^{2} \ .
\end{equation}
The 3-D vector field of the magnetic field is then,
\begin{equation}
        {\bf B_j}({\bf x})
        = B_j \left({d_j \over r_j}\right)^2 {{\bf r}_j \over r_j} \ ,
\end{equation}
Such a magnetic field model with a single buried magnetic (unipolar)
charge can be adequate for a sunspot. For a bipolar active region,
at least two magnetic charges with opposite magnetic polarities are 
necessary.

A general magnetic field can be constructed by superposing the
$n_{mag}$ fields of $j=1,...,n_{mag}$ magnetic charges, defined as,
\begin{equation}
        {\bf B}({\bf x}) = \sum_{j=1}^{n_{\rm mag}} {\bf B}_j({\bf x})
        = \sum_{j=1}^{n_{\rm mag}}  B_j
        \left({d_j \over r_j}\right)^2 {{\bf r_j} \over r_j} \ .
\end{equation}
where the depth $d_j$ of a magnetic charge $j$ is,
\begin{equation}
        d_j = 1-\sqrt{x_j^2+y_j^2+z_j^2} \ ,
\end{equation}
and the distance $r_j$ between the magnetic charge position $(x_j, y_y, z_j)$
and an arbitrary location $(x,y,z)$ where the calculation of a magnetic field
vector is desired, is defined by,
\begin{equation}
        r_j = \sqrt{(x-x_j)^2+(y-y_j)^2+(z-z_j)^2} \ .
\end{equation}
The origin at the spherical coordinate system is chosen at the center
of the Sun, while the sub-photospheric location of a buried magnetic
charge is defined at the position $(x_j, y_j, z_j)$, the vertical
direction ($\theta =0$) is perpendicular to the solar surface,
and the physical unit of length scales (such as $d_j$ in Eq.~15
and $r_j$ in Eq.~16) is in solar radii.  

The multi-pole magnetic field is divergence-free, since
\begin{equation}
        \nabla \cdot {\bf B} = \nabla \cdot (\sum_j {\bf B}_j)
        = \sum_j (\nabla \cdot {\bf B}_j) = 0 \ ,
\end{equation}
while the force-freeness is fulfilled with second-order accuracy
in $(\propto b^2 r^2 \sin^2{ \theta } )$ for the solution
of the vertical-current approximation of Eqs.~(3)-(6)
(Aschwanden 2013a),
\begin{equation}
        \nabla \times {\bf B} =
        \nabla \times \sum_{j=1}^{N_{\rm m}} {\bf B}_j \approx
        \sum_{j=1}^{N_{\rm m}} (\nabla_j \times {\bf B}_j) = 
        \sum_{j=1}^{N_{\rm m}} \alpha_j ({\bf r}) {\bf B}_j =
        \alpha ({\bf r}) {\bf B} \ .
\end{equation}
Numerical tests of comparing the analytical approximation solution
with other nonlinear force-free field codes have been conducted for
a large number of simulated and observed magnetograms, and
satisfactory agreement with other NLFFF codes has been established
(e.g., Aschwanden 2013a, 2013b, 2016; Aschwanden and Malanushenko
2013; Warren et al.~2018).

Note that 4 free parameters are required for a potential field solution
$(x_j, y_j, z_j, B_j)$, and 5 free parameters for a nonpotential
solution, i.e., $(x_j, y_j, z_j, B_j, b_j)$, for every magnetic
unipolar charge $j$. A key parameter for 
the investigation of field twisting and braiding studied here,
is the number $N_{twist}$ of helical turns of a loop, which can 
be calculated from the length $L_j$ of a magnetic field line (that
represents the axis of a loop), and the associated nonlinear 
parameter $b_j$ (Eq.~11), 
\begin{equation}
	N_{twist}= { b_j\ L_j \over 2 \pi} = {\varphi_j \over 2 \pi} \ ,
\end{equation}
where $\varphi_j$ is the rotation angle of a twisted loop. 
The relationship between the potential field component $B_P$, the
nonpotential field component $B_{NP}$, and the azimuthal magnetic
field component $B_{\varphi}$ is depicted in Fig.~(1c), which defines the
so-called {\sl nonpotentiality angle} $\mu_{NP}$ between the potential and
nonpotential field direction. The Pythagorean theorem 
of a rectangular triangle yields then,
\begin{equation}
	B_{NP}^2 = B_P^2 + B_{\varphi}^2		\ ,
\end{equation}
which together with the magnetic energy definition $E=B^2/8\pi$ is 
consistent with the definition of the free energy
$E_{free}=E_{\varphi}$ (Aschwanden 2013b),
\begin{equation}
	E_{NP} = E_P + E_{\varphi} = E_P + E_{free} \ .
\end{equation}
We can now calculate the nonpotentiality angle $\mu_{NP}$
from the energy ratio $(E_P/E_{NP})$, or from the mean magnetic 
field ratio $(B_P/B_{NP})$ (see geometrical relationship in Fig.~1c),
\begin{equation}
	 \mu_{NP} = \arccos{ \left( \sqrt{E_P \over E_{NP}} \right) } =
	 	\arccos{ \left( {B_P \over B_{NP}} \right) } \ .
\end{equation}
While the rotation angle $\varphi=2 \pi N_{twist}$ (Eq.~19) 
produces the helical twist of loops, the resulting nonpotentiality 
angle $\mu_{NP}$ between potential and nonpotential field lines can 
be expressed as (see geometrical relationship in Fig.~1c),
\begin{equation}
	\mu_{NP} = \arctan{ \left( {\varphi r \over L} \right) }
	    = \arctan{ \left( 2 \pi N_{twist} { r \over L } \right) } \ ,
\end{equation}
where $r$ is the radius of the azimuthal field strength
$B_{\varphi}(r)$ dependence dropping to about half of its peak value. 
Theoretical calculations yield
$r/L \approx 0.1-0.4$ (Fig.~4 in Aschwanden 2013a), which
yields an estimated ratio of $\mu_{NP}/\varphi \approx (0.1-0.4)$
and can be tested with the alternative definition of Eq.~(22).
Equating Eqs.~(22) and (23) yields the relationship between
magnetic field ratio $(B_P/B_{NP})$ and the twist angle 
$\varphi r/L$, according to the geometrical relationship shown in 
Fig.~1c. 

The main {\sl caveat} of the VCA-NLFFF code is the limited accuracy
of magnetic field modeling for large helical twist numbers (say
for $N_{twist} \gapprox 2$). A pathological case is the
Gold-Hoyle flux rope (Gold and Hoyle 1960), which has been
modeled for twist numbers $N_{twist}=0, 1, 2, 3, 4, 5$ (Aschwanden
2013a, Appendix A). It reveals unrealistic distortions for large
twist numbers $N_{twist} \gapprox 2$, and thus indicates a breakdown
of the approximation made in the VCA-NLFFF code, for such extreme cases.
This limitation, however, does not affect the relatively high
accuracy of the active regions analyzed here, which all consistently
yield a low value of $N_{twist} \approx 0.1-0.2$. Moreover, the
Gold-Hoyle flux rope exceeds the kink instability criterion, and
thus may be not relevant for stable loops. 

\subsection{	Braiding Topology and Linkage Number 		}

While the previous section quantifies helically twisted structures,
we turn now to braided (coronal) structures.
The topology of braiding can be characterized by the so-called
{\sl Gauss linkage (or linking) number} $N_{link}$, which is 
defined as follows:
{\sl The linkage number is a topological parameter describing
two curves that does not change as the curves are distorted 
without crossing through each other} (e.g., Priest 2014).
Each curve is given a direction, and reversing one of
the directions changes the linkage number $N_{link}$ by
$(-1)$. Each crossing has a sign $(+1)$ or $(-1)$, depending on
whether the first curve is in front or behind the other,
and the linkage number is half the sum of the signed
crossings. Related to the linkage number is the magnetic
helicity (for definitions see, e.g., Priest 2014).
We will just use the linkage number here, which represents
the number of times that each curve winds around the other (Fig.~2). 
If the two linked curves are closed, the linkage number is
an integer number, while it can be generalized to a
non-integer number for open curves (e.g., D\'emoulin et al.~2006
and references therein).

A braiding process is clearly recognizable when two curvi-linear 
structures exhibit at least 1-3 turns around each other (Fig.~2), 
so we will require a minimum linkage number of $|N_{link}| \ge 1$
for the identification of a braiding process, which is a strict
lower limit for closed curves (since a non-zero integer number
is required). Magnetic field lines in the solar corona are
generally open curves, whose braiding can be characterized 
by non-integer numbers. If this linkage number
in closed curves is less than unity, we will call it
{\sl helically twisted loops}, rather than {\sl braided loops}.
Fortunately, our parameterization of a nonlinear force-free
magnetic field solution provides the number of helical twists,
$N_{twist}$ (Eq.~19), which we can directly compare with the linkage 
number $N_{link}$ of braiding. Four examples are visualized in
Fig.~(2), showing two parallel loop strands with no braiding (which
has a Gauss linkage number of $L=0$), a helically twisted loop
with $L=1$, and braided loop strands with linkage numbers of
$L=2$ and $L=3$. Each loop can be distorted (e.g. by fitting
a loop geometry to an observed magnetic field model), but does 
not change its linkage number $L$ as long the number of mutual 
crossing points is not changed. In the four examples shown in
Fig.~(2) we have chosen integers for the linkage numbers, but it
is trivial to adjust for non-integer numbers in loop pairs 
with open curves.

\subsection{	Kink and Torus Instability 			}

Testing the (circular and helical) loop structures
observed in the solar corona, we have to address their 
stability. Circular loops generally indicate a near-potential
magnetic field, which tends to be stable, while sigmoid
structures indicate a nonpotential magnetic field, 
possibly being unstable. Since the potential
field solution represents the (magnetic) energy minimum 
configuration, we would expect that potential fields are perfectly
stable, while non-ponential fields have free energy that can be
dissipated and thus the loops are not necessarily stable. 
On the other side, the
magnetic field solution that we derived with the vertical-current
approximation method (Eqs.~7-11) is divergence-free and force-free
(to second-order accuracy), and thus the force-freeness 
indicates a stable or a marginally stable condition. 

However, it was recognized early on that
force-free flux tubes can be prone to the helical kink instability
(e.g., see textbook by Priest 2014, p.266), 
or to the torus instability (e.g., Kliem and T\"or\"ok 2006; 
Isenberg and Forbes 2007),
if the helical twist exceeds about one turn. The detailed instability
criterion depends on additional conditions, such as the axial
magnetic field, the plasma pressure gradient, magnetic field line
curvature, and line-tying. Hood and Priest (1979) calculated
the kink instability criterion for a force-free field and 
uniform twist to be $3.3\pi$, which corresponds to $N_{twist}
=3.3 \pi / 2 \pi = 1.65$ turns, while this value was found to
vary in a typical range of $N_{twist} \approx 1-3$ for other
magnetic field configurations. Further calculations including
the line-tying effect of the dense photosphere yielded a
threshold of $2.5 \pi$ or $N_{twist}=2.5 \pi / 2 \pi = 1.25$
turns (Hood and Priest 1981).

The evolution of twisting coronal magnetic flux tubes with
a compressible zero-beta, ideal MHD code has been simulated,
which showed the formation of a helically twisted flux tube
that evolved along a sequence of force-free equilibria with
progressively stronger helical twist. The system becomes unstable
for a critical twist of $2.5 \pi \le {\Phi}_c \le 2.75 \pi$, which
corresponds to $N_{twist}= 1.25-1.38 $ turns (T\"or\"ok and 
Kliem 2003). 

A similar process observed in laboratory plasmas (in tokamaks and
spheromacs) is the torus instability, an expansion 
instability of a toroidal current ring, for which an equilibrium 
was established by Shafranov, (called the Kruskal-Shafranov limit
$k \ge - \Phi/L$). Qualitative agreement between
spheromak experiments and solar coronal mass ejections was
pointed out by Kliem and T\"or\"ok (2006).
Further MHD simulations of an embedded flux rope showed
the formation of an aneurism-like structure when it erupts,
which was identified as torus instability
(Isenberg and Forbes 2007).
MHD simulations of magnetic reconnection in a kinking flux rope
were also performed with high initial twist ($\Phi \gapprox 6\pi$),
which corresponds to $N_{twist}=6 \pi/2\pi =3 $ helical turns
(Kliem et al.~2010), but such a configuration with multiple
helical turns has never been reported from solar observations
in non-flaring conditions. There was one single observation 
reported of a multiple-turn spiral flux tube with about 4
turns (Gary and Moore 2004), 
which occurred during an eruption from an active region,
but its morphology could equally well be interpreted as a 
(quasi-periodic) chain of multiple vortices of the 
Kelvin-Helmholtz instability (Aschwanden 2019, Section 11.8). 

In summary, there is strong evidence that loop structures
in the force-free corona do not exhibit helical twisting
with more than one turn, while multiple helical turns,
which would be required for a coronal braiding process,
are inherently unstable to the kink or torus instability,
as demonstrated by analytical calculations and numerical
MHD simulations. 

\section{	DATA ANALYSIS		 		}

\subsection{	Active Regions Observed with SDO 	}

We analyze a data set of 15 active regions observed with
AIA/SDO and HMI/SDO, which is identical to the selection
made in recent studies by Warren et al.~(2012, 2018).
The first study (Warren et al.~2012) contains a 
{\sl differential emission measure (DEM)} analysis of 
EIS/Hinode data in the temperature range of 
$T_e \approx 1-10$ MK. The second study (Warren
et al.~2018) compares three different magnetic field
extrapolation codes. The times and heliographic positions 
of the analyzed 15 active regions are given in Table 1.

\subsection{	Data Analysis Method			}

The basic feature that is unique to our magnetic field modeling
method is the inclusion of the observed geometry of coronal 
loop structures from EUV images, besides the photospheric 
magnetograms $B_z(x,y)$, while the transverse components
$B_x(x,y)$ and $B_y(x,y)$ are ignored due to their larger
uncertainties than the line-of-sight component $B_z(x,y)$, 
partially caused by the non-forcefreeness of the magnetic 
field in the photosphere and lower chromosphere. 
This magnetic field extrapolation code makes use of the
vertical-current approximation of the nonpotential field,
called the VCA-NLFFF code.
A description, performance tests, and measurements of
magnetic free energies is given in Aschwanden (2016), the
theory in Aschwanden (2013a), numerical simulations and tests
in Aschwanden and Malanushenko (2013), the free energy
concept in Aschwanden (2013b), additional applications 
to STEREO data in Sandman et al.~(2009), 
Aschwanden and Sandman (2010), Sandman and Aschwanden (2011), 
Aschwanden et al.~(2012, 2015), Aschwanden (2013c),
comparisons with other NLFFF codes in Aschwanden et al.~(2014a),
Warren et al.~(2018), and applications to IRIS, IBIS, and ROSA data
in Aschwanden (2015) and Aschwanden et al.~(2016).
The VCA-NLFFF code is also publicly available in the SSW
library encoded in the {\sl Interactive Data Language (IDL)}, 
see website {\sl http://www.lmsal.com/
$\sim$aschwand/software/}.

\subsection{	Input Parameters			}

We show an example of a VCA-NLFFF run in Fig.~3, which refers
to event \#14 (listed in Table 1). Each of the 15 analyzed
images encompasses an active region, within a field-of-view of
$FOV=0.47 \pm 0.08 \ R_{\odot}$. The input parameters of event
\#14 (listed on the right side of Fig.~3) includes the instrument
(AIA, with the wavelengths 94, 131, 171, 193, 211, 335 \ang ),
the NOAA active region number 11339, and the heliographic position
at the center of the image, N20W03. The automated loop detection
with the OCCULT-2 code (Aschwanden, De Pontieu and Katrukha 2013)
is controlled by a few tuning parameters, such as the maximum
acceptable misalignment angle ($a_{mis}=20^\circ$), the lowpass
filter (nsm1=1 pixel), the highpass filter (nsm2=nsm1+2=3 pixels),
the maximum number of magnetic source components ($n_{mag}=30$),
the maximum number of detected structures ($N_{struc}=1000$),
the minimum and maximum number of iterations ($n_{itmin}=10$,
$n_{itmax}=50$), the proximity limit between a detected loop 
position and a magnetic charge ($prox_{min} < 10$ magnetic 
source depths), the minimum loop length ($l_{min}=5$ pixels),
the minimum curvature radius ($r_{min}=8$ pixels), the flux
threshold ($q_{thresh1}=0$), and the bipass-filtered flux
threshold ($q_{thresh2}=0$).

The VCA-NLFFF code executes 3 major tasks: (i) the decomposition
of a magnetogram $B_z(x,y)$ into a finite number of magnetic
sources, characterized by their sub-photospheric positions
$(x_j, y_j, z_j), j=1,...,n_{mag}$, and the surface magnetic field 
strength vertically above the buried unipolar magnetic charge 
$(B_j)$; (ii) the automated tracing of curvi-linear structures 
in multiple wavelength bands (here from 6 AIA images in coronal 
wavebands); and (iii) forward-fitting of the nonlinear force-free
$\alpha$-parameters $a_j \approx 2 b_j$, 
which is an optimization algorithm
that minimizes the misalignment angles $\mu$ between the observed
loop directions and the theoretically calculated magnetic field 
lines. 

\subsection{	Output Parameters			}

For event \#14 (Fig.~3) we find a number of $n_{det}=483$ 
automatically detected loop structures, of which $n_{loop}=280$ 
loops have a misalignment angle below the limit of 
$a_{mis}=20^{\circ}$. Larger misangle limits have also been
tested, but yielded similar results, except for an increase of
mis-identifications of curvi-linear features such as moss.
The total magnetic potential field energy
of the entire active region amounts to $E_P=2.5 \times 10^{33}$
erg, and the ratio of the nonpotential to the potential energy
is a factor of $E_{NP}/E_P=1.145$, which implies a free energy
of $\approx 15\%$. Comparing among the 15 active regions listed
in Table 1, we see that event \#14 has the largest nonpotential
energy ratio. From the histogram of misalignment angles (Fig.~3
top right) we see that the 2-D misalignment angle has a median
of $\mu_2=6.8^\circ$, which is measured from the line-of-sight
projected loop coordinates. However, from our 3-D best-fit model
we can derive also a 3-D misalignment angle, which has a value
of $\mu_3=12.4^\circ$. These misalignment angles tell us how
well the theoretical magnetic field model converges to the
observed loop geometries.

\subsection{	Helical Twist Analysis			}

Since the vertical-current approximation concept produces
helically twisted field lines by definition, we can now
determine various parameters of the helical twist. 
From the energy ratio $E_P/E_{NP}$ we can directly
measure the nonpotentiality angle between the potential and
nonpotential field by using Eq.~(22), for which we
find a value of $\mu_{NP} = 20.8^\circ$ (for event \#14), or 
$\mu_{NP}=14.7^\circ\pm 2.8^\circ$ in the statistical
average of all 15 analyzed active regions. These
values are based on the energy ratios that have a
statistical average of $E_{NP}/E_P = 1.07 \pm 0.03$
(Table 1).  

The helical twist can be expressed by the average
rotation angle $\varphi_j = b_j L_j, j=1,...,n_{loop}$ 
(Eq.~19), which depends on the measurement of the 
nonlinear $\alpha$-parameters $a_i \approx 2 b_i,
i=1,..,n_{mag}$ (Eq.~10). However, since the 
location of a detected loop is independent of 
the location of magnetic sources, we have to calculate
the $\alpha$-parameter $a_j, j=1,...,n_{loop}$ from 
the nearest magnetic charge $a_i, i=1,...,n_{mag}$. 
Averaging then all rotation angles $\varphi_j= b_j L_j,
j=1,...,n_{loop}$ we find for event \#14 a mean rotation
angle of $\varphi = 69^\circ \pm 47^\circ$, which
corresponds to a number of twists, $N_{twist}=
\varphi/360^\circ = 0.19 \pm 0.13$, which is a fifth of
a full turn only, although this active region has the
largest helical twist. The average of all 15 active
regions exhibit a mean rotation twist angle of
$\varphi = 49^\circ \pm 11^\circ$, or number of
twist $N_{twist}=0.14 \pm 0.03$ (Table 1). 
The length $L_j$ has been measured along a field line
from the first to the second footpoint for closed loops,
while $L_j$ was limited by the height range 
($L_j \le h_{max}=0.2 R_{\odot}$)
of the computation box for open field lines.

The distributions of the helical twist numbers $N_{twist}$
are shown separately for each of the 15 analyzed active regions
(listed in Table 1), which exhibit that their values are
always much smaller than a half turn turn, i.e., $N_{twist}
\lapprox 0.5$ (Fig.~4), and have a mean of  
$N_{twist} \approx 0.14 \pm 0.03$), which is much less than the 
kink or torus instability criterion ($|N_{kink}| \gapprox 1$), 
as well as far below any loop braiding scenario ($|N_{link}| \ge 1$). 

\section{	DISCUSSION				}

\subsection{	Helical Twist Measurements 		}

It was noticed that the projected geometry of a helically
twisted flux tube is a sigmoid (or S-shape), most prominently
displayed in SXT/Yohkoh images (Rust and Kumar 1996) and in
EIT/SOHO images (Portier-Fozzani et al.~2001). Also it was
known since early on that a flux tube is prone to the kink
instability if the twisting exceeds about a full turn,
$N_{twist} \gapprox 1$. Based on this criterion one expects
stability below this value, and an eruptive behavior above
this value, and thus eruptions can be predicted from the
value of the twisting number during any time evolution. 
Although such a critical parameter is very useful, 
no method has been developed until recently that would
faciliate quantitative measurements of the helical twist.

Portier-Fozzani et al.~(2001) developed a forward-fitting
method that fitted a parameterized helical field line that is
wound around a circular torus, to loop shapes observed
with EIT, using the solar rotation stereoscopy method 
(for a review see Aschwanden 2011),
and tracked the evolution of the twist angle as a function 
of time. The twist evolved from 
$N_{twist}=210^\circ/360^\circ=0.58$
to $N_{twist}=10^\circ/360^\circ=0.03$ 
over the period of one day, which indicates a relaxation
of the nonpotential magnetic field.

Malanushenko et al.~(2011, 2012) carries out forward-fitting
of a linear force-free field to individual loops detected in
EUV images, in order to obtain the nonlinear $\alpha$-parameter
that determines the helical twist number. The time evolution 
of the so-determined magnetic helicity in the corona can then
be compared to that of the magnetic flux helicity across the
photosphere, which were found to be similar.

Thalmann et al.~(2014) measure the evolution of the free 
(magnetic) energy in an active region observed with Hi-C, 
using the Wiegelmann-NLFFF code (Wiegelmann et al.~2006). 
The free energy is calculated from the azimuthal magnetic
field component $B_{\varphi}(t)$, i.e., $E_{\varphi}(t)=
B_{\varphi(t)}^2 V/8\pi$. They find that the overall twist 
of the flux rope increased by about half a turn within
12 minutes, from $N_{twist} \approx 1.0$ to $N_{twist}
\approx 1.5$. These latter values appear to be close to the
kink instability criterion. 

The first method (Portier-Fozzani et al.~2001) represents
a geometrical fit without magnetic field constraints, while
the second (Malanushenko et al.~2011, 2012) and third method 
(Thalmann et al.~2014) are based on forward-fitting of
a {\sl linear force-free field (LFFF)} magnetic field model, 
which constrains the helical twist of a single loop or loop bundle. 
In comparison, our VCA-NLFFF method fits the nonlinear
$\alpha$-parameter for every loop bundle that is associated 
with a unipolar magnetic charge. 
@ In future work, it will be
@ straightforward to study the evolution of each unipolar
@ magnetic charge in an active region separately, and 
@ to predict eruptions in individual unstable zones of an
@ active region. 
A similar partitioning of unipolar magnetic regions,
along with a decomposition into ``spin helicity'' and
``braiding helicity'', has been modeled in Longcope et al.~(2007).

\subsection{	Coronal Braiding 	 		}

Hi-C observations have been claimed to be the first direct
evidence of braided magnetic fields in the solar corona
(Winebarger et al.~2013, 2014; Cirtain et al.~2013; 
Tiwari et al.~2014; Pant et al.~2015). These observations
show nanoflare-like time evolutions in inter-moss loops 
(Winebarger et al.~2013, 2014), finestructures down to
$0.2\arcsec$ scales (Cirtain et al.~2013), external
triggering of coronal subflares in active regions
(Tiwari et al.~2014), and quasi-periodic flows with
velocities of 13-185 km s$^{-1}$ (Pant et al.~2015),
apparently occurring in braided finestructure. However,
no helical twist number or Gauss linkage number 
has been measured in these high(est)-resolution Hi-C 
observations that could provide a quantitative proof 
of the hypothesized braiding topology and geometry. 
Although it has been argued that braided finestructure
are unresolved by current instruments, the recent observations
with Hi-C (with a pixel size of $\approx 0.1\arcsec$) have
convincingly demonstrated that AIA resolves many of the loops 
(Peter et al. 2013; Brooks et al. 2013, 2016; 
Morton and McLaughlin 2013; Aschwanden and Peter 2017).

Although a few apparently braided structures have been
detected in highpass-filtered 193 \ang\ images, numerical 
simulations of braided multi-thread bundles of coronal 
loops produce braiding at a range of scales, but the
observed EUV intensities may not necessarily reveal the
underlying braided topology (Wilmot-Smith et al.~2009;
Pontin et al.~2017), especially once magnetic reconnection
started (Wilmot-Smith et al.~2010, 2011; Yeates et al.~2010;
Pontin et al.~2011).

A model with magnetic helicity condensation, based on the
inverse cascade from small scales to large scales that
is known from turbulent MHD systems,
provides a mechanism that removes most complex fine
structure from the rest of the corona, resulting in smooth
and laminar loops (Antiochos 2013), 
which could be the reason that we do not
see braided field lines in the Quiet Sun
(Antiochos 2013; Kniznik et al.~2017). 

Most of the 3-D MHD simulations of twisted coronal loops with
random footpoint motion focus on the coronal heating rate 
(e.g., Wilmot-Smith et al.~2011; Yeates et al.~2014; 
Pontin and Hornig 2015; Bourdin et al.~2013, 2015; 
Hansteen et al.~2015; Peter 2015; Dahlburg et al.~2016; 
Reale et al.~2016), but the degree of twisting and/or 
braiding (topology), which is our main focus here, is largely 
unconstrained by observations in these MHD simulations.

The braiding topology has also been mimicked with a
self-organized criticality model, which yields the result that
the power law distributions of (avalanche) energies are 
altered by the presence of net helicity (Berger et al.~2009, 2015).
This is consistent with the fact that free (magnetic) energy
depends on the degree of helical twisting (Eqs.~20-21).

\subsection{	Photospheric/Chromospheric Braiding 	}

The viability of a braiding mechanism has been simulated 
with magneto-convection simulations, where the topological
complexity is quantified in terms of the field line winding,
the finite time topological entropy, or passive scalar mixing.
A benchmark test that infers photospheric flows from 
sequences of magnetograms with local correlation tracking
shows that complex tangling of photospheric motions occurs on
a time scale of hours, but no global net winding is induced
(Ritchie et al.~2016; Candelaresi et al.~2018).

Numerical (resistive) MHD simulations of twisted and 
(more complex)
braided magnetic fields show that both fields can relax
to stable force-free equilibria, reaching force-free
end states, but the electric current structures in these
final states differ significantly between the (diffuse)
braided field and the (sigmoidal) twisted field
(Prior and Yeates 2016a, 2016b).

Photospheric magnetic braiding models with Alfv\'enic wave
turbulence, in which the coronal field lines are subject 
to slow random footpoint motion, have been developed by 
van Ballegooijen et al.~(2014, 2017), but the resulting coronal 
heating rate was found to be insufficient to balance the
conductive and radiative losses, given the observed 
velocities ($v \approx 1$ km s$^{-1}$) of the footpoint motions. 

Magnetic field line braiding in the photosphere, chromosphere,
and transition region is fundamentally different from the corona,
because the latter is largely in a stable force-free state (except
during flares or eruptive processes), while the former domains
are located in non-forcefree zones, where turbulent MHD models
are more appropriate than in the force-free corona.  

\section{	CONCLUSIONS 				}

In this study we measure the helical twist number of coronal
field lines, which is identical to the Gauss linkage number
in braiding topologies. Our conclusions are the following:

\begin{enumerate}

\item{The vertical-current approximation nonlinear force-free
field (VCA-NLFFF) code yields a mean nonpotentiality angle
of $\mu_{NP} = 15^\circ \pm 3^\circ$ for a set of 15 active
regions, which is a quantitative measure how much nonpotential
the average magnetic field is compared with a potential field.
The related free energy ratio averages to a value of 
$E_{NP}/E_P=1.07\pm 0.03$, so the analyzed active regions
have a free energy of $E_{free}/E_{P} \approx 4\%-10\%$,
which is consistent with a larger study of 173 events, 
where $E_{free}/E_{P} \approx 1\%-25\%$ was found
(Aschwanden, Xu, and Jing 2014b).}

\item{The mean nonpotentiality angle $\mu_{NP}$ 
corresponds of a mean
rotational twist angle of $\varphi = 49^\circ \pm 11^\circ$,
measured over the length of a field line, which corresponds 
to a relatively low twist number of $N_{twist}=0.14 \pm 0.03$ 
turns. This is the first quantitative result of the helical 
twist number in the solar corona, based on nonlinear force-free
magnetic modeling. A few linear force-free fits have been 
previously pioneered by Malanushenko et al.~(2011, 2012) 
and Thalmann et al.~(2014).}

\item{The helical twist number $N_{twist}$ of a magnetic 
field line is equivalent to the Gauss linkage number $N_{link}$.  
Our result of a relatively small value of the helical twist
number, i.e., $N_{twist} \approx 0.15$, is far below the
kink instability or torus instability value $|N_{twist}|
\approx 1.0$, and thus is consistent with the observed
stability of coronal loops. An absolute upper limit of our 
measurements in (non-flaring) active regions 
is found to be a half turn ($N_{twist} \lapprox 0.5$).}

\item{Numerical 3-D MHD simulations with high linkage
numbers $|N_{link}| \gapprox 1$ are neither consistent
with the observed low twist numbers, nor are they consistent
with the observed stability of coronal loops.}
 
\item{Parker-type nanoflaring scenarios assume braiding
of magnetic field lines, which then trigger magnetic
reconnection above some threshold value of field line
misalighment between adjacent field lines. However, no significant
braiding ($N_{link}=1, 2, ...$) can build up in the 
force-free solar corona according to our observed
helical twist numbers, and thus Parker-type nanoflaring
is more likely to occur in non-forcefree environments,
such as in the chromosphere and transition region, rather
than in the force-free corona. Magneto-convection below
and above the photosphere and the associated MHD turbulence
is a viable driver for chromospheric nanoflaring.}

\end{enumerate}

Future progress in our understanding of coronal heating
mechanisms strongly depends on accurate magnetic field models.
Nonlinear force-free field models (NLFFF) have been criticized
because of the non-forcefreeness of the lower boundary in the
photosphere (DeRosa et al.~2009), but improvements have been
implemented by preprocessing of photospheric fields 
(Wiegelmann et al.~2006), forward-fitting to automatically
traced coronal loops (Aschwanden 2013a), and magneto-static
modeling (Wiegelmann et al.~2017). Nevertheless, the existing
NLFFF codes are still not able to model satisfactorily
the high-resolution data of Hi-C, although a faithfull
attempt has been made (Thalmann et al.~2014). It appears
that many loop structures are resolved in Hi-C images
(Peter et al.~2013; Brooks et al.~2013; Winebarger et al.~2013, 2014;
Morton and McLaughlin 2013; Alexander et al.~2013; Tiwari et al.~2016;
Aschwanden and Peter 2017). On the other side, high-resolution
magnetic field measurements with HMI/SDO do not match up the
resolution of Hi-C, but may be available with DKIST in near future
(Tritschler et al.~2016).

\bigskip
We acknowledge useful discussions with Aad van Ballegooijen and 
attendees of a DKIST 
science planning meeting, held at Newcastle upon Thyne (UK).
This work was partially supported by NASA contracts NNX11A099G,
NNG04EA00C (SDO/AIA), and NNG09FA40C (IRIS). 

\section*{References} % REFERENCES
\def\ref#1{\par\noindent\hangindent1cm {#1}}

\ref{Alexander, C.E., Walsh, R.W., Regnier, S., Cirtain, J., Winebarger, A.R., 
	et al. 2013, ApJ 775, L32.
 	{\sl Anti-parallel EUV Flows Observed along Active Region Filament 
	Threads with Hi-C}}
\ref{Antiochos, S.K. 2013, ApJ 590, 547.
	{\sl Helicity condensation as the origin of coronal
	and solar wind structure}}
\ref{Aschwanden, M.J.: 2004, {\sl Physics of the Solar Corona. 
	An Introduction}, Berlin: Springer and Praxis, p.216.}
\ref{Aschwanden, M.J. and Sandman, A.W. 2010,
	{\sl Bootstrapping the coronal magnetic field with STEREO:
	Unipolar potential field modeling}, ApJ 140, 723.}
\ref{Aschwanden,M.J. 2011, Living Reviews in Solar Physics 8, 5.
 	{\sl Solar stereoscopy and tomography}}
\ref{Aschwanden, M.J., Wuelser,J.-P., Nitta, N.V., et al. 2012,
 	{\sl First 3D Reconstructions of Coronal Loops with the STEREO A 
	and B Spacecraft: IV. Magnetic Field Modeling with Twisted
	Force-Free Fields}, ApJ 756, 124.}
\ref{Aschwanden, M.J. 2013a, SoPh 287, 323, 
        {\sl A nonlinear force-free magnetic field approximation
        suitable for fast forward-fitting to coronal loops. I. Theory}}
\ref{Aschwanden, M.J. 2013b, SoPh 287, 369.
        {\sl A nonlinear force-free magnetic field approximation
        suitable for fast forward-fitting to coronal loops.
        III. The free energy}}
\ref{Aschwanden, M.J. 2013c,
	{\sl Nonlinear force-free magnetic field fitting to coronal loops
	with and without stereoscopy}, ApJ 763, 115.}
\ref{Aschwanden, M.J. and Malanushenko, A. 2013, SoPh 287, 345.
        {\sl A nonlinear force-free magnetic field approximation
        suitable for fast forward-fitting to coronal loops.
        II. Numerical Code and Tests}}
\ref{Aschwanden, M.J., DePontieu, B., and Katrukha, E. 2013,
 	{\sl Optimization of Curvi-Linear Tracing Applied to Solar 
	Physics and Biophysics}, Entropy, 15(8), 3007.}
\ref{Aschwanden, M.J., Sun, X.D., and Liu,Y. 2014a,
 	{\sl The magnetic field of active region 11158 during the 
	2011 February 12-17 flares: Differences between photospheric 
	extrapolation and coronal forward-fitting methods}, ApJ 785, 34.}
\ref{Aschwanden, M.J., Xu, Y., Jing, J. 2014b, ApJ 797, 50.
	{\sl Global energetics of solar flares. I. Magnetic energies}}
\ref{Aschwanden, M.J. 2015,
	{\sl Magnetic energy dissipation during the 
	2014 March 29 solar flare},
	ApJ 804, L20.}
\ref{Aschwanden, M.J., Schrijver, C.J., and Malanushenko, A. 2015,
	{\sl Blind stereoscopy of the coronal magnetic field},
	SoPh 290, 2765.}
\ref{Aschwanden, M.J. 2016, ApJSS 224, 25.
        {\sl The vertical current approximation nonlinear force-free
        field code - Description, performance tests, and measurements
        of magnetic energies dissipated in solar flares}}
\ref{Aschwanden, M.J., Reardon, K., and Jess, D.B. 2016,
	{\sl Tracing the chromospheric and coronal magnetic field
	with AIA, IRIS, IBIS, and ROSA data},
	ApJ 826, 61.}
\ref{Aschwanden, M.J. and Peter, H. 2017, ApJ 840, 4. 
	{\sl The width distribution of loops and strands in the solar
	corona - Are we hitting rock bottom}}
\ref{Aschwanden, M.J. 2019, {\sl New Millennium Solar Physics},
	Section 11.8, New York: Springer, (in press),
	http://www.lmsal.com/$\sim$aschwand/bookmarks$\_$books2.html.}
\ref{Berger, M.A. and Asgari-Targhi, M. 2009, ApJ 705, 347.
	{\sl Self-organized braiding and the structure
	of coronal loops}}
\ref{Berger, M.A., Asgari-Targhi, M., and DeLuca, E.E. 2015,
	J.~Plasma Phys. 81/4, 395810404.
	{\sl Self-organized braiding in solar coronal loops}}
\ref{Bourdin, P.A., Bingert, S., and Peter, H. 2013,
	A\&A 555, A123.	
	{\sl Observationally driven 3D MHD model of the
	solar corona above an active region}}
\ref{Bourdin, P.A., Bingert, S., and Peter, H. 2015,
	A\&A 580, A72.
	{\sl Coronal energy input and dissipation in a solar
	active region 3D MHD model}}
\ref{Boyd, T.J.M., Sanderson, J.J.: 2003, {\sl The Physics of Plasmas},
        Cambridge University Press, Cambridge, p.102.}
\ref{Brooks, D.H., Warren, H.P., Ugarte-Urra, I., and Winebarger, A.R.
 	2013, ApJ 772, L19.
 	{\sl High Spatial Resolution Observations of Loops in the Solar Corona}}
\ref{Brooks, D.H., Reep, J. W., and Warren, H.P. 2016,
 	ApJ 826, L18.
 	{\sl Properties and Modeling of Unresolved Fine Structure Loops 
	Observed in the Solar Transition Region by IRIS}}
\ref{Candelaresi, S., Pontin, D.I., Yeates, A.R., et al. 2018,
	ApJ 864, 157. 
	{\sl Estimating the rate of field line braiding in the
	solar corona by photospheric flows}}
\ref{Cirtain, J.W., Golub, L., Winebarger, A.R. 2013,
	{\sl Energy release in the solar corona from spatially
	resolved magnetic braids}.}
\ref{Dahlburg, R.B., Einaudi, G., Taylor, B.D., et al. 2016,
	ApJ 817, 47.
	{\sl Observational signatures of coronal loop heating
	and cooling driven by footpoint shuffling}}
\ref{D\'emoulin, P., Pariat, E., and Berger, M.A. 2006, 
	{\sl Basic properties of mutual magnetic helicity},
	SoPh 233, 3.} 					
\ref{DeRosa, M.L., Schrijver, C.J., Barnes, G., et al. 2009, ApJ 696, 1780.
 	{\sl A critical assessment of nonlinear force-free field modeling of 
	the solar corona for active region 10953.}}
\ref{Gary, G.A. and Moore, R.L. 2004, ApJ 611, 545.
	{\sl Eruption of a Multiple-Turn Helical Magnetic Flux Tube 
	in a Large Flare: Evidence for External and Internal 
	Reconnection That Fits the Breakout Model of Solar Magnetic Eruptions}}
\ref{Gold, T. and Hoyle, R. 1960, MNRAS 120, 89.
	{\sl On the origin of solar flares}}
\ref{Hansteen, V., Guerreiro, N., de Pontieu, B., et al.~2015,
	ApJ 811, 106.
	{\sl Numerical simulations of coronal heating through
	footpoint braiding}}
\ref{Hassanin, A. and Kliem, B. 2016, ApJ 832, 106.
	{\sl Helical kink instability in a confined solar eruption}}
\ref{Hood, A.W. and Priest, E.R. 1979, {\sl Kink instability of solar
	coronal loops as the cuase of solar flares}, SoPh 64, 303.}
\ref{Hood, A.W. and Priest, E.R. 1981, {\sl Critical conditions for
	magnetic instabilities in force-free coronal loops},
	Geophys. and Astrophysical Fluid Dynamics 17(3-4), 297.}
\ref{Huang, Z., Xia, L., Nelson, C.J., et al. 2018, ApJ 854, 80.
	{\sl Magnetic braids in eruptions of a spiral structure
	in the solar atmosphere}}
\ref{Isenberg, P.A. and Forbes, T.G. 2007,
	{\sl A 3-D line-tied magnetic field model for solar eruption},
	ApJ 670, 1453.}
\ref{Kliem, B. and T\"or\"ok, T. 2006,
	{\sl Torus instability}, Phys.Rev.Lett. 96, 255002.} 
\ref{Kliem, B. et al. 2010, {\sl Reconnection of a kinking flux rope
	triggering the ejection of a microwave and hard X-ray source.
	II. Numerical modeling}, SoPh 266, 91.}
\ref{Kniznik, K.J., Antiochos, S.K., and DeVore, C.R. 2017,
	ApJ 835, 85.
	{\sl The role of magnetic helicity in structuring the
	solar corona}}
\ref{Koleva, K., Madjarska, M.S., Duchlev, P., et al. 2012, 
	A\&A 540, A127.
	{\sl Kinematics and helicity evolution of a loop-like
	eruptive prominence}}
\ref{Kumar, P., Srivastava, A.K., Filippov, B., and Uddin, W. 2010,
	SoPh 266, 39.
	{\sl Multiwavelength study of the M8.9/3B solar flare
	from AR NOAA 10960}}
\ref{Kumar, P., Cho, K.S., Bong, S.C., et al. 2012,
	{\sl Initiation of coronal mass ejection and associated
	flare caused by helical kink instability observed by SDO/AIA}} 
\ref{Liu, Y. and Schuck, P.W. 2012, ApJ 761, 105.
	{\sl Magnetic energy and helicity in two emerging
	active regions in the Sun}}
\ref{Longcope, D.W., Ravindra, B., and Barnes, G. 2007.
	ApJ 668, 571.
	{\sl Determining the source of coronal helicity through
	measurements of braiding and spin helicity fluxes
	in active regions}}
\ref{Malanushenko, A., Yusuf, M.H., and Longcope, D.W. 2011, ApJ 736, 97.
	{\sl Direct measurements of magnetic twist in the solar corona}}
\ref{Malanushenko, A., Schrijver, C.J., DeRosa, M.L., Wheatland, M.S.,
	and Gilchrist, S.A. 2012, ApJ 756, 153.
	{\sl Guiding nonlinear force-free modeling using coronal
	observations: First results using a quasi-grad-Rubin scheme}}
\ref{Morton, R.J., and McLaughlin, J.A. 2013.
 	A\&A 553, L10.
 	{\sl Hi-C and AIA observations of transverse magnetohydrodynamic 
	waves in active regions}}
\ref{Pant, V., Datta, A., and Banerjee, D. 2015,
	ApJ 801, L2.
	{\sl Flows and waves in braided solar coronal magnetic
	structures}}
\ref{Parker, E.N. 1988, ApJ 330, 474,
	{\sl Nanoflares and the solar X-ray corona}}
\ref{Peter, H. 2015,
	Phil.Trans.Royal Soc. A 373/2042, 20150055.
	{\sl What can large-scale MHD numerical experiments
	tell us about coronal heating?}} 
\ref{Peter, H., Bingert, S., Klimchuk, J.A., de Forest, C., Cirtain, J.W., 
	Golub, L., Winebarger, A. R., Kobayashi, K., and Korreck, K.E.
 	2013, A\&A 556, 104.
 	{\sl Structure of solar coronal loops: from miniature to large-scale}}
\ref{Pontin, D.I., Wilmot-Smith, A.L., Hornig, G., et al. 2011,
	A\&A 525, A57.
	{\sl Dynamics of braided coronal loops. II. Cascade
	to multiple small-scale reconnection events}}
\ref{Pontin, D.I. and Hornig, G. 2015, ApJ 805, 47.
	{\sl The structure of current layers and degree of
	field-line braiding in coronal loops}}
\ref{Pontin, D.I., Janvier, M., Tiwari, S.K., Galsgaard, K., et al.~2017,
	{\sl Observational signatures of energy release in braided coronal loops}}
\ref{Portier-Fozzani, F., Aschwanden, M., D\'emoulin, P. et al. 2001,
	{\sl Measurement of coronal magnetic twist during loop
	emergence of NOAA 8069}}
\ref{Priest, E.R.: 1982, {\sl Solar Magnetohydrodynamics},
        Geophysics and Astrophysics Monographs Volume 21,
        D. Reidel Publishing Company, Dordrecht, p.125.}
\ref{Priest, E. 2014,				
        {\sl Magnetohydrodynamics of the Sun},
        Cambridge, UK: Cambridge University Press, p.111.}	
\ref{Prior, C. and Berger, M.A. 2012, SoPh 278, 323.
	{\sl On the shape of force-free field lines in the
	solar corona}}
\ref{Prior, C. and Yeates, A.R. 2016a, A\&A 587, A125.
	{\sl Twisted versus braided magnetic flux ropes in
	coronal geometry. I. Construction and relaxation}}
\ref{Prior, C. and Yeates, A.R. 2016b, A\&A 591, A16.
	{\sl Twisted versus braided magnetic flux ropes in
	coronal geometry. II. Comparative behaviour}}
\ref{Raouafi, N.E. 2009, ApJ 691, L128.
	{\sl Observational evidence for coronal twisted flux rope}}
\ref{Reale, F., Orlando, S., Guarrasi, M., et al.~2016,
	ApJ 830, 21.
	{\sl 3-D MHD modeling of twisted coronal loops}}
\ref{Ritchie, M.L., Wilmot-Smith, A.L., and Hornig, 2016,
	{\sl The dependence of coronal loop heating on the
	characteristics of slow photospheric motions}}
\ref{Rust, D.M. and Kumar, A. 1996, ApJ 464, L199.
	{\sl Evidence for helically kinked magnetic flux ropes
	in solar eruptions}}
\ref{Sandman, A.W., Aschwanden, M.J., DeRosa, M.L., et al. 2009,
	{\sl Comparison of STEREO/EUVI loops with potential magnetic
	field models}, SoPh 259, 1.}
\ref{Sandman, A.W. and Aschwanden, M.J. 2011,
	{\sl A new method for modeling the coronal magnetic field with
	STEREO and submerged dipoles}, SoPh 270, 503.}
\ref{Sturrock, P.A.: 1994, {\it Plasma Physics. -- An Introduction to the
        Theory of Astrophysical, Geophysical and Laboratory Plasmas},
        Cambridge University Press, Cambridge, p.216.}
\ref{Thalmann, J.K., Tiwari, S.K., and Wiegelmann, T. 2014, ApJ 780, 102,
	{\sl Force-free field modeling of twist and braiding-induced
	magnetic energy in an active region corona}}
\ref{Tiwari, S.K., Alexander, C.E., Winebarger, A.R., et al. 2014,
	{\sl Trigger mechanism of solar subflares in a braided
	coronal magnetic structure}}
\ref{Tiwari, S.K., Moore, R.L., Winebarger, A.R., and Alpert, S.E.
 	2016, ApJ 816, 92.
 	{\sl Transition-region/Coronal Signatures and Magnetic Setting 
	of Sunspot Penumbral Jets: Hinode (SOT/FG), Hi-C, and SDO/AIA 
	Observations}}
\ref{T\"or\"ok, T. and Kliem, B. 2003,
	{\sl The evolution of twisting coronal magnetic flux tubes},
	A\&A 406, 1043.}
\ref{T\"or\"ok, T., Berger, M.A., and Kliem, B. 2010, A\&A 516, A49.
	{\sl The writhe of helical structures in the solar corona}}
\ref{Tritschler, A., Rimmele, T.T., Berukoff, S., et al. 2016,
        Astron.Nachrichten 337/10, 1064.
        {\sl Daniel K. Inouye Solar Telescope: High-resolution
        of the dynamic Sun}}
\ref{van Ballegooijen, A.A., Asgari-Targhi, M., and Berger, M.A. 2014,
	{\sl On the relationship between photospheric footpoint
	motions and coronal heating in solar active regions}}
\ref{van Ballegooijen, A.A., Asgari-Targhi, M., and Voss, A. 2017,
	ApJ 849, 46. 
	{\sl The heating of solar coronal loops by Alfv\'en wave
	turbulence}}
\ref{Warren, H.P., Winebarger, A.R., and Brooks, D.H. 2012,
	{\sl A systematic survey of high-temperature emission
	in solar active regions}, ApJ 759, 141.}
\ref{Warren, H.P., Crump, N.A., Ugarte-Urra, I., Sun, X.,
        Aschwanden, M.J., and Wiegelmann, T. 2018, ApJ 860, 46.
        {\sl Toward a quantitative comparison of magnetic field
        extrapolations and observed coronal loops}}
\ref{Wedemeyer-B\"ohm, S., Scullion, E., Steiner, O., et al. 2012,
	Nature 486, Issue 7404, 505.
	{\sl Magnetic tornadoes as energy channels into the solar corona}}
\ref{Wiegelmann T., Inhester, B., and Sakurai, T. 2006, 
	SoPh 233, 315.
	{\sl Preprocessing of Vector Magnetograph Data for a Nonlinear 
	Force-Free Magnetic Field Reconstruction}}
\ref{Wiegelmann, T., Neukirch, T., Nickeler, D.H., Solanki, S.K., Barthol, et al.
 	2017, ApJSS 229, 18.
	{\sl Magneto-static Modeling from Sunrise/IMaX: Application to an 
	Active Region Observed with Sunrise II}}
\ref{Wilmot-Smith, A.L., Hornig, G., and Pontin, D.I. 2009,
	ApJ 704, 1288.
	{\sl Magnetic braiding and quasi-separatrix layers}}
\ref{Wilmot-Smith, A.L., Pontin, D.I., and Hornig, G. 2010,
	A\&A 516, A5.
	{\sl Dynamics of braided coronal loops}}
\ref{Wilmot-Smith, A.L., Pontin, D.I., Yeates, A.R., et al.~2011,
	A\&A 536, A67.
	{\sl Heating of braided coronal loops}}
\ref{Winebarger, A.R., Walsh, R.W., Moore, R., et al. 2013,
	ApJ 771, 21.
	{\sl Detecting nanoflare heating events in subarcsecond
	inter-moss loops using Hi-C}}
\ref{Winebarger, A.R., Cirtain, J., Golub, L., and DeLuca, E. 2014,
	ApJ 787, L10.
	{\sl Discovery of finely structured dynamic solar corona
	observed in the Hi-C telescope}}
\ref{Yeates, A.R., Hornig, G., and Wilmot-Smith, A.L. 2010,
	Phys.Rev.Lett. 105/8, 085002.
	{\sl Topological constraints on magnetic relaxation}}
\ref{Yeates, A.R., Bianchi, F., Welsch, B.T., et al.~2014,
	A\&A 564, A131.
	{\sl The coronal energy input from magnetic braiding}}
%@@@REFERENCES

%%%%%%%%%%%%%%%%%%%%%%%%%%%%%% TABLES  %%%%%%%%%%%%%%%%%%%%%%%%%%%%%%%%%%

\begin{table}
\tabletypesize{\normalsize}
%\tabletypesize{\footnotesize}
\caption{Data selection parameters (observation date and time,
heliographic position of center of image, field-of-view of image
in units of solar radii), and data analysis results
from AIA/SDO and HMI/SDO data: 
ratio of nonpotential to potential energy ($E_{NP}/E_P$),
nonpotentiality angle between nonpotential and potential field $\mu$, 
number of helical twist turns ($N_{twist}$),
helical twist rotation angle $\varphi$,
based on the VCA-NLFFF forward-fitting code.}
\medskip
\begin{tabular}{llllllll}
\hline
\#&
Observation&
Heliographic&
Field&
Energy&
Nonpot.&
Number&
Rotation \\
&
date and time&
position&
of view&
ratio&
angle&
of twist&
angle\\
&
(UT)&
&
FOV ($R_{\sun}$)&
$E_{NP}/E_P$&
$\mu_{NP}$ (deg)&
$N_{twist}$&
$\varphi$ (deg)\\
\hline
 1 & 20100619 012742 & N29E19 & 0.335 &  1.044 & 11.9$^\circ$ &  0.10$\pm$ 0.06 &    34$^\circ\pm19^\circ$ \\
 2 & 20100621 011627 & N26W07 & 0.408 &  1.039 & 11.2$^\circ$ &  0.11$\pm$ 0.09 &    40$^\circ\pm33^\circ$ \\
 3 & 20100723 143256 & S27E23 & 0.419 &  1.058 & 13.6$^\circ$ &  0.14$\pm$ 0.09 &    49$^\circ\pm33^\circ$ \\
 4 & 20100929 232134 & N21W21 & 0.481 &  1.064 & 14.2$^\circ$ &  0.18$\pm$ 0.12 &    63$^\circ\pm42^\circ$ \\
 5 & 20110121 134056 & N21E04 & 0.523 &  1.085 & 16.3$^\circ$ &  0.12$\pm$ 0.12 &    42$^\circ\pm43^\circ$ \\
 6 & 20110131 105511 & S24E35 & 0.408 &  1.061 & 13.8$^\circ$ &  0.14$\pm$ 0.10 &    51$^\circ\pm37^\circ$ \\
 7 & 20110212 150157 & S21E19 & 0.367 &  1.034 & 10.4$^\circ$ &  0.08$\pm$ 0.07 &    29$^\circ\pm24^\circ$ \\
 8 & 20110411 113035 & N17E34 & 0.533 &  1.087 & 16.4$^\circ$ &  0.17$\pm$ 0.11 &    61$^\circ\pm41^\circ$ \\
 9 & 20110415 004705 & N19W11 & 0.512 &  1.091 & 16.8$^\circ$ &  0.17$\pm$ 0.13 &    61$^\circ\pm47^\circ$ \\
10 & 20110419 130206 & N15W05 & 0.512 &  1.078 & 15.6$^\circ$ &  0.13$\pm$ 0.09 &    45$^\circ\pm32^\circ$ \\
11 & 20110702 030812 & N15E22 & 0.388 &  1.087 & 16.5$^\circ$ &  0.16$\pm$ 0.10 &    57$^\circ\pm34^\circ$ \\
12 & 20110725 090557 & N25W15 & 0.575 &  1.047 & 12.2$^\circ$ &  0.14$\pm$ 0.12 &    48$^\circ\pm41^\circ$ \\
13 & 20110821 115609 & N15E05 & 0.575 &  1.055 & 13.2$^\circ$ &  0.12$\pm$ 0.10 &    42$^\circ\pm34^\circ$ \\
14 & 20111108 184444 & N20W03 & 0.575 &  1.145 & 20.8$^\circ$ &  0.19$\pm$ 0.13 &    69$^\circ\pm47^\circ$ \\
15 & 20111110 110329 & N21W26 & 0.502 &  1.096 & 17.2$^\circ$ &  0.13$\pm$ 0.11 &    48$^\circ\pm38^\circ$ \\
   &                 &        &       &        &      &                 &                 \\
mean & & & 0.47$\pm$0.08 & 1.07$\pm$ 0.03 & 14.7$^\circ\pm2.8^\circ$ & 0.14$\pm$0.03 & 49$^\circ\pm11^\circ$ \\
\hline
\end{tabular}
\end{table}

%%%%%%%%%%%%%%%%%%%%%%%%%%%%%% FIGURES %%%%%%%%%%%%%%%%%%%%%%%%%%%%%%%%%%

\begin{figure}
\centerline{\includegraphics[width=0.8\textwidth]{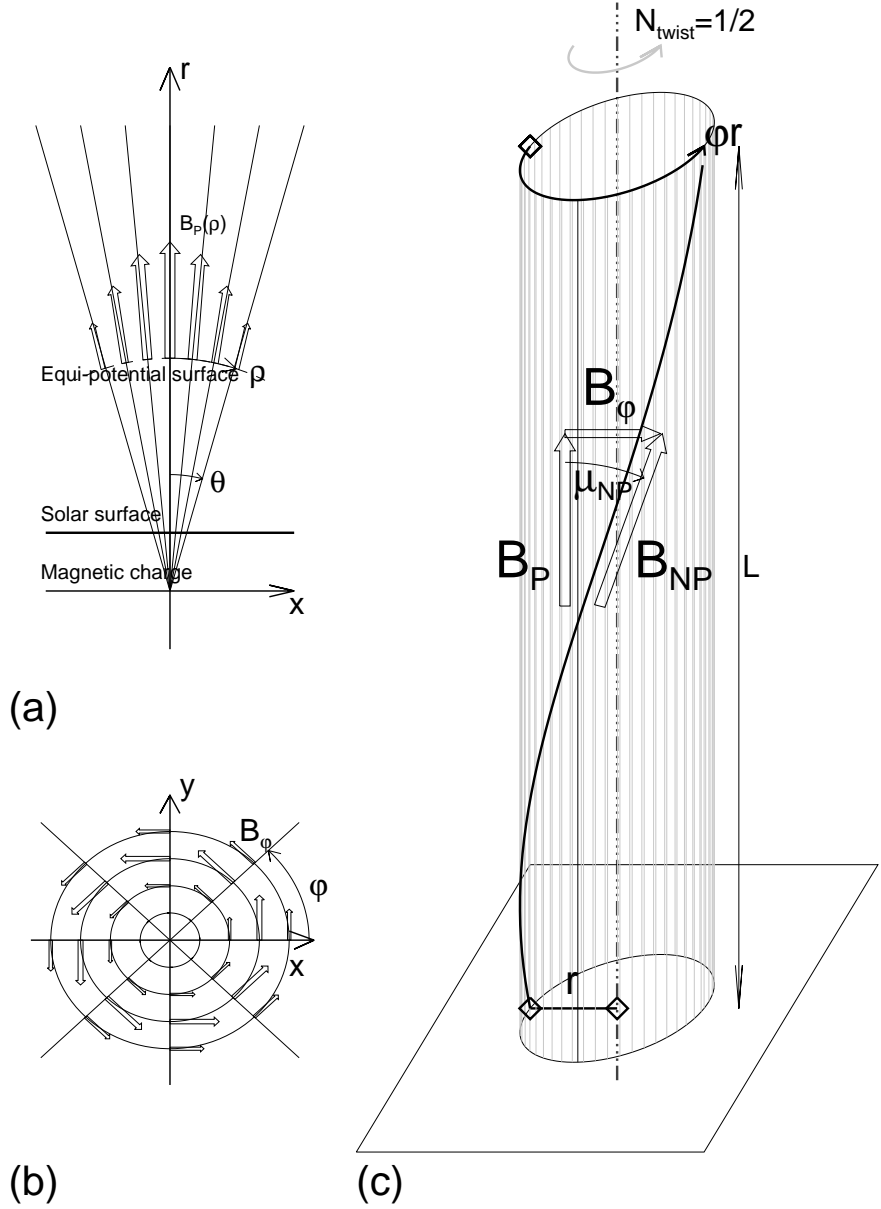}}
\caption{Geometric concept of vertical current approximation model:
(a) A magnetic charge is buried below the solar surface, with radial
field lines $B_r$ pointing away with inclination angles $\theta$; (b) 
Top view onto x-y plane with azimuthal field component $B_{\varphi}$
indicated; (c) The relationship of the radial potential field $B_r=B_P$,
the nonpotential field $B_{NP}$, and the azimuthal field component
$B_{\varphi}$, which are distorted by a nonpotentiality angle 
$\mu_{NP}$ in a helically
twisted flux tube. The twist corresponds to a half turn ($N_{twist}=0.5$)
over a length $L$.}
\end{figure}

\begin{figure}
\centerline{\includegraphics[width=1.0\textwidth]{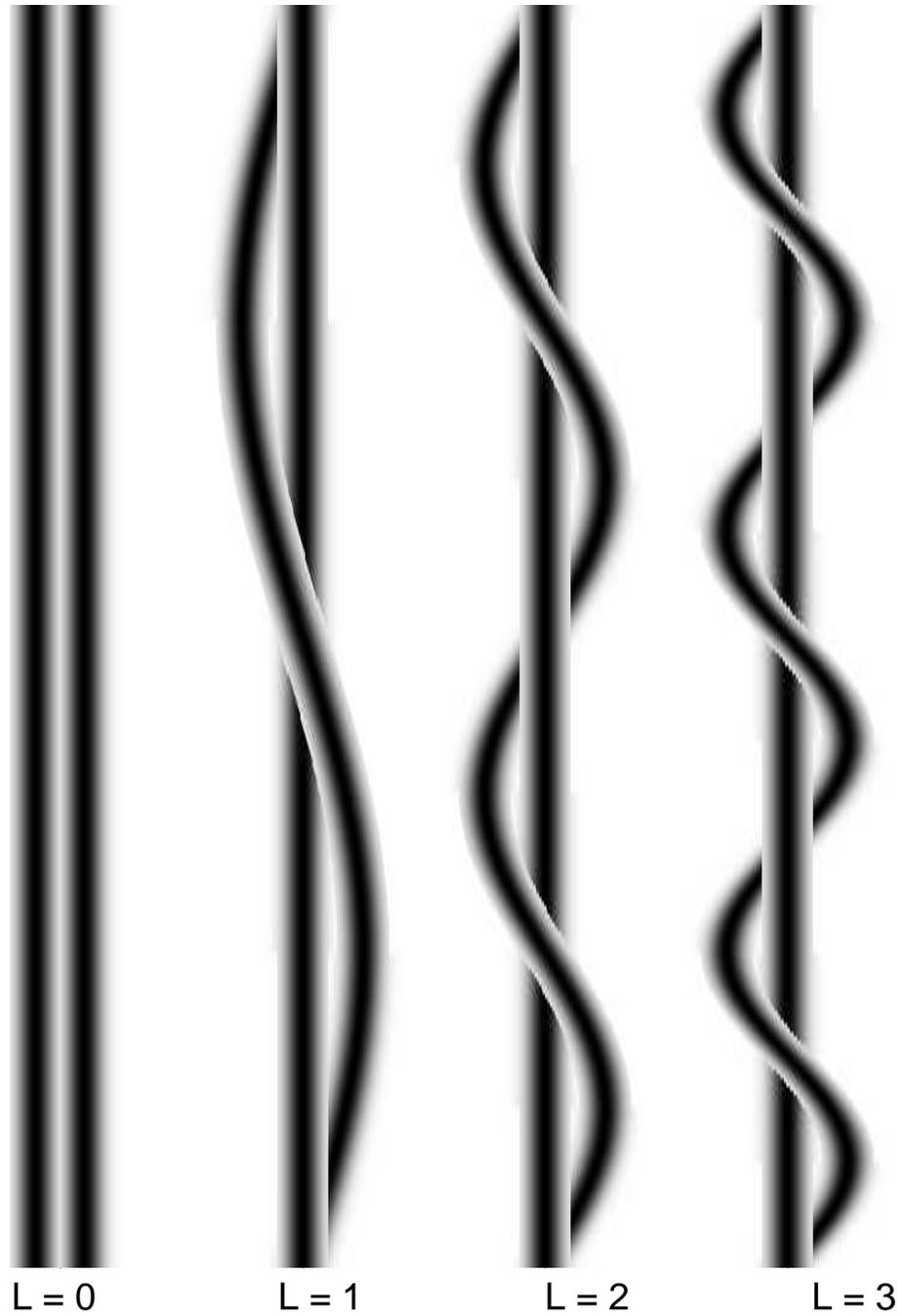}}
\caption{Four pairs of loops with different Linkage numbers (L=0, 1, 2, 3).
The untwisted case $L=0$ corresponds to a potential field model,
$L=1$ to a twisted loop near the torus instability, and two cases
of braided loops ($L=2, 3$) that are unstable.}
\end{figure}

\begin{figure}
\centerline{\includegraphics[width=0.9\textwidth]{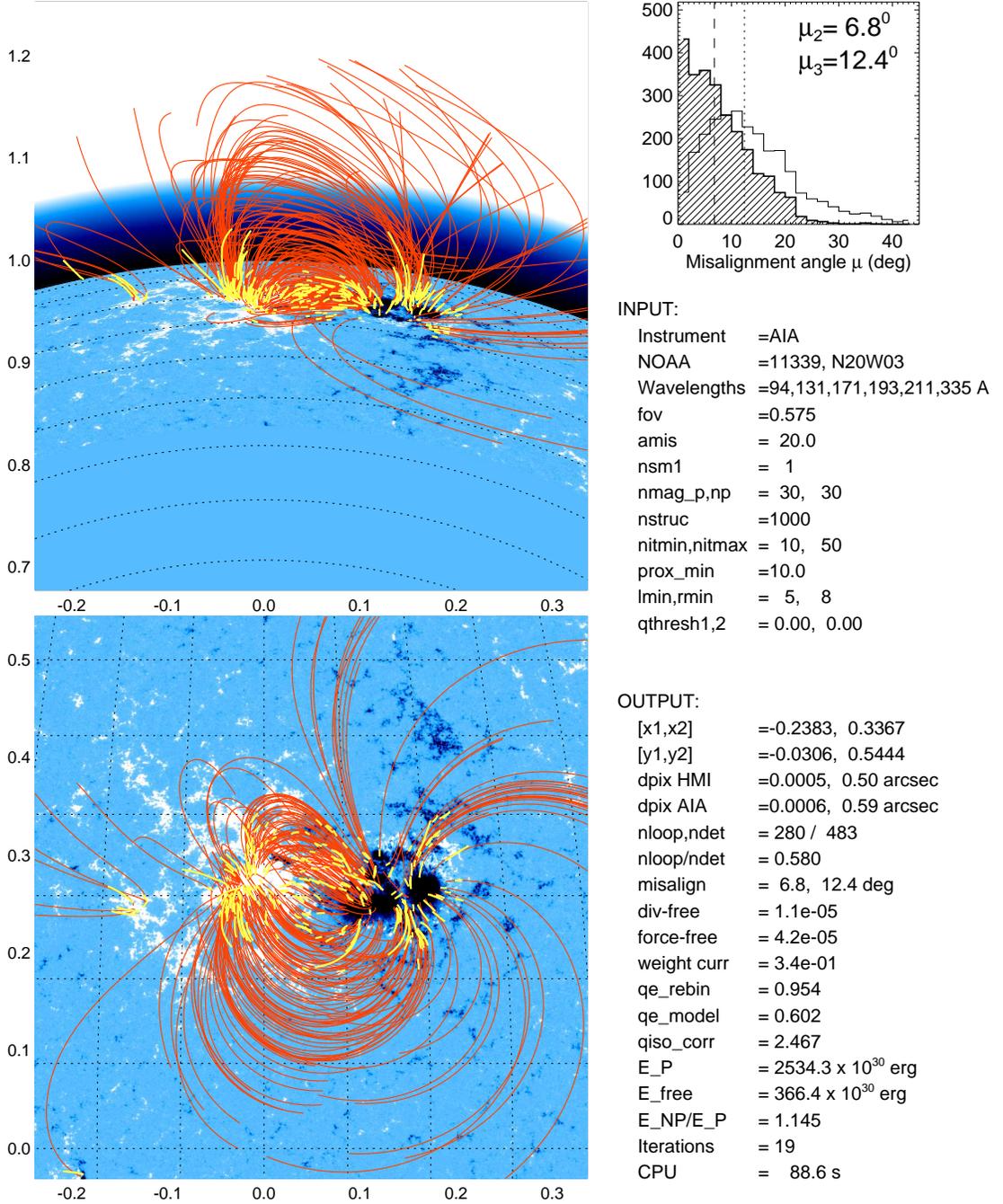}}
\caption{Magnetic field model obtained from forward-fitting of the VCA-NLFFF
code to AIA/SDO and HMI/SDO data of event \#14, observed on
2011 November 08, 18:44:44 UT, with heliographic position N20W03 at 
the image center. {\sl Bottom left:} Automatically traced loops
in AIA/SDO in all coronal wavelength bands (yellow), with the
best fitting magnetic field lines (red curves) overlayed on
the HMI magnetogram (blue). {\sl Top left:} 3-D magnetic field
configuration rotated by $90^0$ as it would be seen from a top
view from far above the solar north pole. {\sl Top right:}
Histogram of 2-D ($\mu_2$) and 3-D ($\mu_3$) misalignment angles
between the observed loop directions (yellow curves) and the 
theoretical magnetic field model (red curves), each measured in
7 segments of the 280 traced loops.}
\end{figure}

\begin{figure}
\centerline{\includegraphics[width=1.0\textwidth]{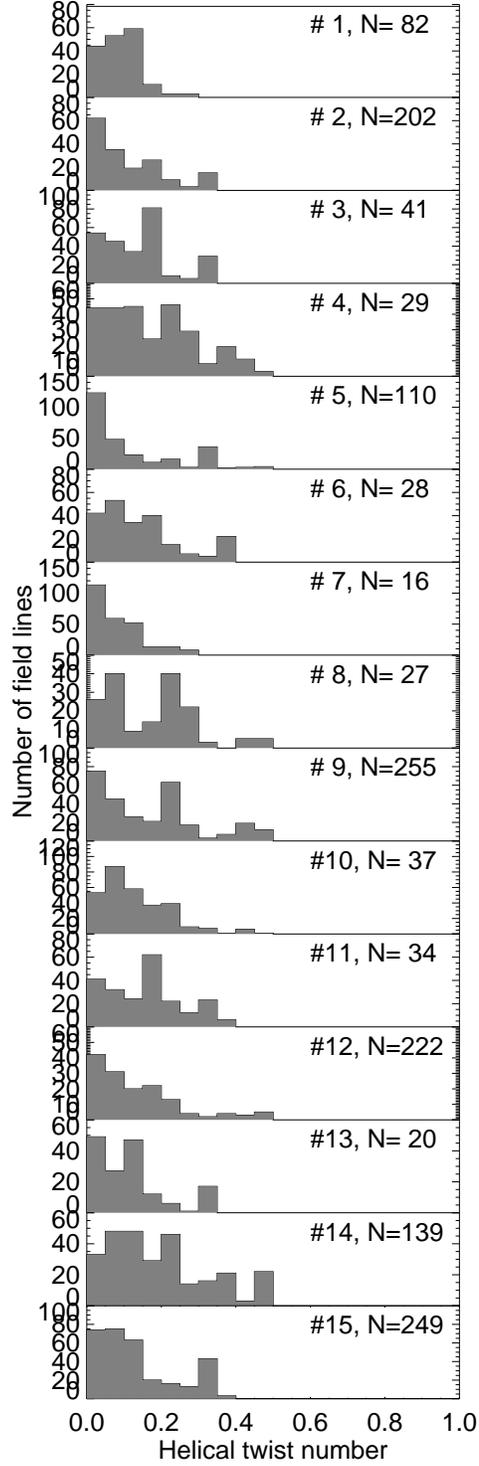}}
\caption{Statistical distributions of helical twist numbers $N_{twist}$
shown separately for each of the 15 active regions listed in Table 1.
The number of measured field lines is given in the top right corner.
Note that the helical twist number never exceeds a value of 
a half turn (i.e., $N_{twist} \lapprox 0.5$).}
\end{figure}

\end{document}